\documentclass[aps,prb,twocolumn,groupedaddress,floats,showpacs]{revtex4}
\usepackage{latexsym}
\usepackage{dcolumn}
\usepackage[dvips]{graphicx}
\usepackage{amssymb}
\usepackage{graphics}
\usepackage{amsmath}
\usepackage{epsf}

\renewcommand{\vec}[1]{\mbox{\boldmath$#1$}}

\newcommand{\up}{\uparrow}
\newcommand{\down}{\downarrow}

\newcommand{\ve}{\vec e}
\newcommand{\vJ}{\vec J}
\newcommand{\vm}{\vec m}

\newcommand{\vn}{\vec n}

\newcommand{\vmu}{\mbox{\boldmath $\mu$}}
\newcommand{\vtau}{\mbox{\boldmath $\tau$}}

\begin{document}

\title{Theory of the spin-torque-driven ferromagnetic resonance \\in a
ferromagnet/normal-metal/ferromagnet structure}
\author{Joern N.\ Kupferschmidt, Shaffique Adam, and Piet W.\ Brouwer}

\affiliation{Laboratory of Atomic and Solid State Physics,
Cornell University, Ithaca, NY 14853-2501, USA}

\date{\today}

\begin{abstract}
We present a theoretical analysis of current driven ferromagnetic 
resonance in a ferromagnet--normal-metal--ferromagnet tri-layer. This
method of driving ferromagnetic resonance was recently realized
experimentally by
Tulapurkar {\em et al.} [Nature {\bf 438}, 339 (2005)] and Sankey {\em
et  al.} [Phys.\ Rev.\ Lett.\ {\bf 96}, 227601 (2006)]. The precessing
magnetization rectifies the alternating current applied to drive the
ferromagnetic resonance and leads to the generation of a dc
voltage. Our analysis shows that a second mechanism to generate a dc
voltage, rectification of spin currents emitted by the precessing
magnetization, has a contribution to the dc voltage that is of
approximately  equal size for the thin ferromagnetic films used in the
experiment.
\end{abstract}   

\pacs{76.50.+g,72.25.Ba,75.75.+a,85.75.-d}

\maketitle

\section{Introduction}

A decade ago, Slonczewski\cite{kn:slonczewski1996} and
Berger\cite{kn:berger1996} predicted that a spin-polarized current
passing through a ferromagnet exerts a torque on its magnetic moment.
The past decade has shown an abundance of experiments that have
confirmed this theoretical prediction.\cite{kn:tsoi1998,kn:sun1999,
kn:wegrowe1999,kn:myers1999,kn:katine2000} Since spin-polarized
currents are easily generated by passing an electrical current through
a ferromagnet, the `spin transfer torque' opens the way for all-electrical
manipulation of nanoscale magnetic
devices.\cite{kn:tserkovnyak2005,kn:brataas2006}

Very recently, two groups have been able to use the spin
torque to drive and detect ferromagnetic resonance in a
ferromagnet--normal-metal--ferromagnet (FNF)
trilayer.\cite{kn:tulapurkar2005,kn:sankey2006} These experiments are
designed such that the magnetization direction of one of the 
ferromagnets is fixed by anisotropy forces, whereas the other magnet
is made of a softer ferromagnetic material or has a more symmetric
shape so that its magnetization can more easily respond to the applied
current or to an applied magnetic field. 
In both experiments, an alternating
electrical current is used to drive
the ferromagnetic resonance, whereas the magnetization precession is
detected through the dc voltage generated by rectification of the
applied ac current by the time-dependent resistance of the
device.\cite{kn:tulapurkar2005,kn:sankey2006} 
The theoretical analysis of this 
experimental setup is the subject of this article.

Not only does a spin-polarized current have an effect on the
direction of the magnetization of a ferromagnet, a 
time varying magnetization also causes the flow of spin
currents in normal metal conductors in electrical contact
to the ferromagnet. This `spin
emission' was proposed by Tserkovnyak, Brataas, and
Bauer as the cause of enhanced damping of 
ferromagnetic resonance in
thin ferromagnetic films in good electrical contact to a normal metal
substrate.\cite{kn:tserkovnyak2002} It is also the mechanism
underlying Berger's earlier prediction of the excitation of a dc voltage
by a precessing magnetization in an unbiased FNF 
trilayer\cite{kn:berger1999} 
(see also Refs.\ \onlinecite{kn:azevedo2005,kn:wang2006}).

Spin emission affects the experiments of Refs.\
\onlinecite{kn:tulapurkar2005} and \onlinecite{kn:sankey2006} in two
different ways. First, through the enhancement of the damping spin
emission broadens the ferromagnetic resonance. Second, the free
layer's precessing magnetization emits alternating spin currents,
which, in turn, generate a dc voltage through the time-varying 
spin-dependent conductance of the free layer.\cite{kn:berger1999} 
That way, spin emission 
provides an alternative to rectification of the applied ac current
as a mechanism for the generation of a dc voltage in 
these experiments. Our calculations show that both consequences of 
spin emission appear or disappear together: If spin emission gives a
significant contribution to the damping of the ferromagnetic resonance
--- which is the case for the few nm-thick free-layer ferromagnets used in 
the experiments ---, then it also provides a sizable contribution to
the measured dc voltage, and vice versa.

In the remainder of this article we present the detailed theory of the
electrical-current driven ferromagnetic resonance needed to arrive at
the above conclusion. In addition, our theory allows us to calculate
how the ferromagnetic resonance frequency, the resonance width, and
the asymmetry of the resonance lineshape are affected by
embedding the free ferromagnetic layer into the FNF
trilayer. Our calculation proceeds in three parts. In Sec.\ \ref{sec:2}
we derive general expressions for the spin transfer torque, 
which we then apply to the calculation of the
magnetization motion in Sec.\ \ref{sec:3}. The generated dc voltage is
calculated in Sec.\ \ref{sec:4}. We conclude in Sec.\ \ref{sec:5}.

\section{Spin transfer torque}
\label{sec:2}

\begin{figure}
\epsfxsize=0.7\hsize
\hspace{0.01\hsize}
\epsffile{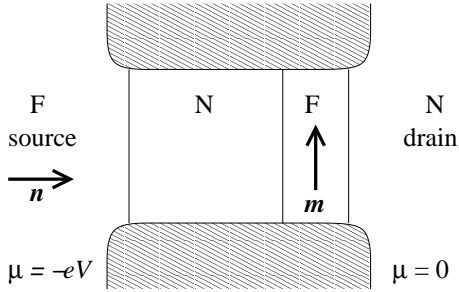}
\caption{\label{fig:1} Schematic drawing of the
ferromagnet--normal-metal--ferromagnet trilayer considered here. The
left ferromagnet, with magnetization direction $\vn$, acts as the
source reservoir. The right ferromagnet is the free layer. Its
magnetization direction $\vm$ can change in response to the applied
current. The currents $J(+)$ and $J(-)$ in the text are evaluated at 
the right and left sides of the free ferromagnetic layer, respectively.}
\end{figure}

A schematic drawing of the system we consider is shown in Fig.\
\ref{fig:1}. It consists of a ferromagnetic source reservoir, held at
electric voltage $V$, a thin normal-metal spacer layer, a thin
ferromagnetic layer, and a normal-metal drain reservoir. The direction
$\vn$ of the magnetization in the ferromagnetic source is considered
to be fixed, whereas the direction $\vm$ of the magnetization in the
thin layer can change under the influence of an electrical current or
an applied magnetic field.

The nonequilibrium
spin transfer torque arises from the discontinuity of the spin
current $\vJ_{\rm s}$ across the free ferromagnetic
layer,\cite{kn:slonczewski1996,kn:waintal2000,kn:stiles2002}
\begin{equation}
  \vtau_{\rm ne} = -[\vJ_{\rm s}(+) - \vJ_{\rm s}(-)],
  \label{eq:torque}
\end{equation}
where $\vJ_{\rm s}(+)$ and $\vJ_{\rm s}(-)$ are spin currents at the
normal-metal--ferromagnet interfaces measured on the side of the drain 
reservoir and the spacer, respectively.
We take the spacer layer and the free ferromagnet are 
sufficiently thin, so that all voltage drops occur across the
ferromagnet--normal-metal interfaces, and spin 
relaxation can be neglected.\cite{endnote15} 
(Note that neglecting spin relaxation 
in the spacer layer is justified for the $10$ nm thick Cu spacer used in 
the experiment of
Ref.\ \onlinecite{kn:sankey2006}, which has a thickness
much below the spin 
diffusion length in Cu. The spin diffusion length $l_{\rm sf}$
in ferromagnets can be
much smaller, however, and the experiments of Refs.\ \onlinecite{kn:tulapurkar2005,kn:sankey2006}
have a free layer
thickness $d$ comparable to $l_{\rm sf}$, not $d \ll l_{\rm sf}$.
Still, we do not expect a strong effect of spin-flip scattering in
this case, since the spin accumulation in the free layer remains fixed 
collinear with the direction $\vm$ of the magnetic moment, whereas the
driving and detection of the ferromagnetic resonance depend on the 
misalignment of the two magnetic moments in the device.\cite{endnote16})
With these assumptions, the charge
currents $J_{\rm c}(\pm)$ and the spin currents $\vJ_{\rm s}(\pm)$
can be expressed directly in terms of the charge and spin 
accumulations $\mu_{\rm c}$ and $\vmu_{\rm s}$ in the spacer
layer. For the charge and spin currents $J_{\rm c}(-)$ and $J_{\rm
  s}(-)$ one has two sets of equations, one arising from the interface
with the ferromagnetic source
reservoir,\cite{kn:brataas2000b,kn:tserkovnyak2005,kn:brataas2006}
\begin{eqnarray}
  J_{\rm c}(-) &=& \frac{1}{e} \left[2 G_+ (\mu_{\rm c} + e V) + 2 G_-
  \vmu_{\rm s} \cdot \vn \right], \nonumber \\
  \vJ_{\rm s}(-) &=& - \frac{\hbar}{2 e^2}
  \left[2 G_+ \vmu_{\rm s} \cdot \vn + 2 G_- (\mu_{\rm c} + e
  V)\right] \vn \nonumber \\ && \mbox{}
  + \frac{\hbar}{2 e^2} \left[ 2 G_1 (\vmu_{\rm s} \times \vn) \times
  \vn + 2 G_2 \vmu_{\rm s} \times \vn \right], ~~~
  \label{eq:jn}
\end{eqnarray}
and one arising from the interface with the free ferromagnetic
layer,\cite{kn:tserkovnyak2002,kn:tserkovnyak2005,kn:brataas2006}
\begin{eqnarray}
  J_{\rm c}(-) &=& - \frac{1}{e}
  \left[ g_+ \mu_{\rm c} + g_- \vmu_{\rm s} \cdot \vm \right]
  \nonumber \\
  \vJ_{\rm s}(-) &=& \frac{\hbar}{2 e^2}
  \left[ g_- \mu_{\rm c} + g_+ \vmu_{\rm s} \cdot \vm \right] \vm
  \nonumber \\ && \mbox{}
  - \frac{\hbar}{2 e^2} g_1 \left[2 \vmu_{\rm s} \times \vm
  + \hbar \dot{\vm} \right] \times \vm
  \nonumber \\ && \mbox{}
  - \frac{\hbar}{2 e^2} g_2 \left[2 \vmu_{\rm s} \times \vm
  + \hbar \dot{\vm} \right].
  \label{eq:jm}
\end{eqnarray} 
Here $G_{\pm} = (G_{\up} \pm G_{\down})/2$ and $G_1 + i G_2 =
G_{\up\down}$ are determined by the interface conductances for
majority and minority electrons and by the mixing
conductance for the interface between the ferromagnetic source and the
normal-metal spacer, whereas $g_{\pm} = (g_{\up} \pm g_{\down})/2$ and
$g_1 + i g_2 = g_{\up \down}$ represent the equivalent quantities for
the interface between spacer layer and the free ferromagnet and for
the interface
between the free ferromagnet and the source. Numerical
values for these conductance coefficients have been obtained for
the interfaces of
various combinations of ferromagnetic and normal-metal
materials.\cite{kn:xia2002} 

The two sets of equations are slightly different because there are
two ferromagnet--normal-metal interfaces between the spacer layer and
the drain reservoir, whereas there is only one interface between the
spacer layer and the source reservoir, see Fig.\
\ref{fig:1}.\cite{kn:adam2006}
Also, in Eq.\ (\ref{eq:jn}), we
omitted terms proportional to the time derivative $\dot{\vn}$ because
the magnetization of the source reservoir is held fixed. Similarly,
for $J_{\rm c}(+)$ and $\vJ_{\rm s}(+)$ we find
\begin{eqnarray}
  \label{eq:jmm}
  J_{\rm c}(+) &=& - \frac{1}{e}
  \left[ g_+ \mu_{\rm c} + g_- \vmu_{\rm s} \cdot \vm \right]
  \nonumber \\
  \vJ_{\rm s}(+) &=& \frac{\hbar}{2 e^2}
  \left[ g_- \mu_{\rm c} + g_+ \vmu_{\rm s} \cdot \vm \right] \vm
  \nonumber \\ && \mbox{}
  + \frac{\hbar^2}{2 e^2} g_1 \dot{\vm}\times \vm
  + \frac{\hbar^2}{2 e^2} g_2 \dot{\vm}.
\end{eqnarray} 
Note that the charge current $J_{\rm c}$ and the component $\vJ_{\rm
  s} \cdot \vm$ of the spin
current parallel to the direction of the magnetization of the
free layer are conserved. 

The mixing conductances $G_1 + i G_2$ and $g_1 + i g_2$ describe 
the coherent reflection of electrons with spin not collinear with the magnetization directions $\vn$ and $\vm$ off the interface with the fixed and free ferromagnetic layers, respectively. We omitted terms that represent the coherent transmission of electrons with spin not collinear with $\vn$ and $\vm$. The effect of coherent transmission is small for ferromagnets much thicker than the ferromagnetic coherence length, which is usually on the order of only a couple of atomic layers. We refer to Refs.\ \onlinecite{kn:waintal2000,kn:tserkovnyak2002} for a theory in which these processes are included.
Since the imaginary parts $G_2$ and $g_2$ of the mixing conductances
are numerically small for metallic junctions
(20\% or less of $G_1$ and
$g_1$),\cite{kn:xia2002,kn:zwierzycki2005,kn:zimmler2004} 
we set $G_2$ and $g_2$ to zero in the following
calculations. At the end of Sec.\ \ref{sec:4} we discuss how 
our results are modified for finite $G_2$ and $g_2$.

The flow of electrical current through the FNF trilayer generates a
spin-transfer torque only if the magnetization
directions $\vn$ and $\vm$ are not collinear. In the experiment of
Refs.\ \onlinecite{kn:tulapurkar2005} and
\onlinecite{kn:sankey2006} this is achieved by an applied
magnetic field which orients the free-layer magnetization $\vm$ 
at a finite angle with respect to the fixed-layer magnetization 
direction $\vn$ in 
the absence of a current. Following Ref.\
\onlinecite{kn:sankey2006}, we take this angle to be $90$
degrees.
We choose a right-handed set of coordinate 
axes ($\ve_1,\ve_2,\ve_3$) such that $\vn$ points along $\ve_1$ and 
$\vm$ points along $\ve_3$ if no current is applied. The application
of a current will cause $\vm$ to deviate from $\ve_3$. We'll be
interested in the linear response regime, in which the magnetization 
components $m_1$ and $m_2$ are proportional to the applied current
$J$.

With an alternating
current bias, $J_{\rm c} = J(t) = \mbox{Re}\, J_0 e^{i \omega t}$, 
Eqs.\ (\ref{eq:jn})
and (\ref{eq:jm}) give five independent equations, from which one can
solve for the five unknown variables, which are
the charge and spin accumulations $\mu_{\rm c}$ and $\vmu_{\rm s}$ in 
the spacer layer and the bias voltage $V$. Solving these to lowest 
order in the applied current, we find that two relevant components
of the spin transfer torque (\ref{eq:torque}) are
\begin{eqnarray}
  \tau_{{\rm ne},1} &=&
  -\frac{\hbar}{2 e}
  \left[J \frac{G_-}{G_{1+}}
  + \frac{\hbar \dot{m}_2 g_1}{e}
  \left(2 - \frac{G_+}{G_{1+}} \right) \right], \\
  \tau_{{\rm ne},2} &=&
  \frac{\hbar}{2 e}
  \frac{\hbar \dot{m}_1 g_1}{e} \left(2 - \frac{g_1}{g_1+G_1}
  \right),
\end{eqnarray}
where we abbreviated
\begin{equation}
  G_{1+} = G_+ + (G_+^2-G_-^2)/g_1.
\end{equation}

\section{Magnetization dynamics}
\label{sec:3}

The magnetization is driven out of equilibrium
by the spin transfer torque of Eq.\ (\ref{eq:torque}). In order to
solve for the full time-dependence of the magnetization, we use the
Landau-Lifshitz-Gilbert equation,\cite{kn:lifshitz1980,kn:gilbert2004}
\begin{eqnarray}
  \dot{\vm} =
  \alpha \vm \times \dot{\vm}
  + \frac{\gamma}{M d}(\vtau_{\rm eq} + \vtau_{\rm ne}).
\end{eqnarray}
Here $M$ is the magnetization per unit length, $d$ is the thickness of
the free ferromagnetic layer, $\gamma$ is the gyromagnetic
ratio, and $\alpha$ is the phenomenological Gilbert damping
parameter. 
The equilibrium torque $\vtau_{\rm eq}$ is the combination of
the torque applied by the external magnetic field and the anisotropy
torque intrinsic to the ferromagnet. Since we are interested in small
deviations from equilibrium, we can expand $\vtau_{\rm eq}$ around the
equilibrium direction $\vm = \ve_3$,
\begin{equation}
  \vtau_{{\rm eq}} = -\frac{M d}{\gamma}
  (\omega_1 m_1' \ve_1' + \omega_2 m_2' \ve_2')
  \times \vm,
\end{equation}
where the frequencies $\omega_1$ and $\omega_2$ are set by the energy cost for
magnetization deviations along principal axes $\ve_1'$ and $\ve_2'$
perpendicular to $\ve_3$. The constants $\omega_1$ and $\omega_2$ depend on the
dipolar field of the pinned layer, the demagnetization field and
coercivity of the free layer, and the applied magnetic
field.\cite{kn:ohandley2000}
The
geometric mean $(\omega_1 \omega_2)^{1/2}$ is the free layer's ferromagnetic
resonance frequency in the absence of electrical contact to the
normal-metal spacer layer and the drain reservoir, whereas
$(\omega_1/\omega_2)^{1/2}$ is the ratio of semi-major and semi-minor axis of
the ellipsoidal magnetization precession in that case.
If $\vtau_{\rm eq}$ is dominated by the
applied magnetic field $H$, one has $\omega_1 = \omega_2 = \gamma H$. 
Rotating to the coordinate system with unit
vectors $\ve_1$ and $\ve_2$, the two components of $\vtau_{\rm eq}$
can be written
\begin{eqnarray}
  \tau_{{\rm eq},1} &=&
  \frac{M d}{\gamma} \left[
  - m_2 (\omega_+ + \omega_- \cos \phi) +
  m_1 \omega_- \sin \phi \right], \nonumber \\
  \tau_{{\rm eq},2} &=& \frac{M d}{\gamma}
  \left[ m_1 (\omega_+ - \omega_- \cos \phi) - m_2 \omega_- \sin \phi \right],
\end{eqnarray}
where $\omega_{\pm} = (\omega_2 \pm \omega_1)/2$ and $\phi/2$ is the rotation angle
between $\ve_1'$ and $\ve_1$.

With an applied ac current, $J(t) = \mbox{Re}\, J_0 e^{i \omega t}$ we
can then solve for the magnetization components $m_1(t) =  \mbox{Re}\,
m_{10} e^{i \omega t}$ and $m_2(t) =  \mbox{Re}\,
m_{20} e^{i \omega t}$, with the result
\begin{eqnarray}
  m_{10} &=& m_0 \frac{(J_0/e) (i \omega + \omega_- \sin \phi)}
  {f(\omega)}, \label{eq:m10} \\
  m_{20} &=& m_0 \frac{(J_0/e)
  (\omega_+ - \omega_- \cos \phi + i\omega (\tilde \alpha_{+} + \tilde \alpha_{-}))}
  {f(\omega)}, ~~~~ \label{eq:m20}
\end{eqnarray}
where we abbreviated
\begin{eqnarray}
  f(\omega) &=& (1 + \tilde \alpha_{+}^2 - \tilde \alpha_{-}^2) \omega^2 -
  2 i \omega( \tilde \alpha_{+} \omega_+ + \tilde \alpha_{-} \omega_- \cos \phi)
  \nonumber \\ && \mbox{}
  + \omega_-^2-\omega_+^2, \\
  m_0 &=& \gamma \hbar G_-/2 d M G_{1+},
\end{eqnarray}
and
\begin{eqnarray}
  \tilde \alpha_{+} &=& \alpha + \frac{g_1 \gamma \hbar^2}{4 d e^2 M}
  \left(4 - \frac{G_+}{G_{1+}} - \frac{g_1}{g_1+G_1} \right),
  \label{eq:alphadef} \\
  \tilde \alpha_{-} &=& \frac{g_1 \gamma \hbar^2}{4 d e^2 M}
  \left(\frac{G_+}{G_{1+}} -\frac{g_1}{g_1+G_1} \right).
\end{eqnarray}
The non-negative
dimensionless numbers $\tilde \alpha_{\pm}$ are the effective Gilbert
damping parameters.\cite{kn:tserkovnyak2002} We need two damping
parameters rather than one since the effective damping is 
anisotropic because of the presence of the second ferromagnet.

\section{DC voltage}
\label{sec:4}

Since we are interested in the dc voltage generated by the applied ac
current, we need to calculate the voltage $V(t)$ to second order in
$J(t)$. This implies that we need to solve Eqs.\ (\ref{eq:jn}) and
(\ref{eq:jm}) to first order in $m_1$ and $m_2$,
\begin{eqnarray}
  V &=&
  J \left( \frac{2 G_1 + g_+}{G_1 g_{1+}}
  + \frac{G_+ + g_1}{2 g_1 G_{1+}} \right)
  - \frac{\hbar \dot{m}_2 G_-}{2 e G_{1+}}
  \nonumber \\ && \mbox{} - \frac{2 (g_1+G_1) g_- G_- m_1}{G_1 g_1
  g_{1+}
  G_{1+}}
  \left( J
  - \frac{e \dot{m}_2 \tilde \alpha_-}{m_0}  \right),~~~
  \label{eq:V}
\end{eqnarray}
where we abbreviated
\begin{equation}
  g_{1+} = 2 g_+ + (g_+^2 - g_-^2)/G_1.
\end{equation}

The two terms in the first line of Eq.\ (\ref{eq:V}), 
which are proportional to
$J$ and $\dot{m}_2$, give an alternating contribution to $V$ only. The
term proportional to $J$ is the dc resistance of the device, whereas
the term proportional to $\dot{m}_2$ is the magnetic contribution to
the admittance. (Electronic contributions to the admittance occur at
higher frequencies than the ferromagnetic resonance frequency and are
not considered in our theory.) 
The dc voltage
follows from the sub-leading terms in the second line of Eq.\
(\ref{eq:V}), which are proportional to $J m_1$
and $ \dot{m_2} m_1$. The contribution proportional to $J m_1$ is
rectification of the applied alternating current by the time-dependent
conductance of the device. The contribution proportional to $\dot{m}_2
m_1$ follows from spin emission by the precessing magnetization of the
free ferromagnet.

The two terms contributing to the dc voltage are easily calculated
using the results of the previous section. Using Eqs.\ (\ref{eq:m10})
and (\ref{eq:m20}), one calculates the averages of the products 
$J m_1$ and $\dot{m}_2 m_1$ over one period of the applied current,
\begin{eqnarray}
  \langle J m_1 \rangle &=&
  \frac{m_0 |J_0|^2}{2 e |f(\omega)|^2}\,
  [\omega\, \mbox{Im}\, f(\omega) + 
  \omega_- \sin \phi\, \mbox{Re}\, f(\omega)],
  \nonumber \\
  \langle \dot{m}_2 m_1 \rangle &=&
  \frac{m_0^2 |J_0|^2 \omega^2
  }{2 e^2 |f(\omega)|^2}
  \label{eq:m2m1}
  \\ && \mbox{} \times
  [\omega_+ - \omega_- \cos \phi - (\tilde \alpha_{+} + \tilde \alpha_{-})
  \omega_- \sin \phi]. \nonumber 
\end{eqnarray}
The dc voltage then follows from substitution into Eq.\ (\ref{eq:V}),
\begin{widetext}
\begin{eqnarray}
  V &=& 
  \frac{m_0 |J_0|^2 g_- G_-(g_1 + G_1)}
  {e G_1 g_1 g_{1+} G_{1+} |f(\omega)|^2}
  \nonumber \\ && \mbox{} \times
  \left\{ \omega^2 (2 \omega_+ \tilde \alpha_+ +
  \omega_+ \tilde \alpha_- + \omega_-  \tilde
  \alpha_- \cos \phi)
  - \omega_- 
  [(1 + \tilde \alpha_+^2 + \tilde \alpha_+ \tilde \alpha_-) 
  \omega^2 + \omega_-^2 - \omega_+^2] \sin \phi
  \right\}.
  \label{eq:Vdc}
\end{eqnarray}
In the limit $\tilde \alpha_{\pm} \ll 1$ (which is appropriate for
most experiments), Eq.\ (\ref{eq:Vdc})
simplifies to the asymmetric Lorentzian
\begin{eqnarray}
  V &=& V_0 \frac{\omega_0^2 -
  (\omega-\omega_0) \delta'}{(\omega-\omega_0)^2 + \delta^2},
  \label{eq:VdcLor}
\end{eqnarray}
with
\begin{eqnarray}
  V_0 &=& \frac{m_0 |J_0|^2  g_- G_-(g_1 + G_1)}
  {4 \omega_0^2 e G_1 g_1 g_{1+} G_{1+}}
  (2 \omega_+ \tilde \alpha_+ +
  \omega_+ \tilde \alpha_- + \omega_-  \tilde \alpha_- \cos \phi),
\end{eqnarray}
\end{widetext}
and
\begin{eqnarray}
  \omega_0^2 &=& \omega_+^2 - \omega_-^2, \nonumber \\
  \delta &=& \tilde \alpha_+ \omega_+ + \tilde \alpha_- \omega_- \cos \phi, 
  \nonumber \\
  \delta' &=& \frac{2 \omega_0 \omega_- \sin \phi}{
  2 \omega_+ \tilde \alpha_+ +
  \omega_+ \tilde \alpha_- + \omega_-  \tilde \alpha_- \cos \phi}.
\end{eqnarray}

The asymmetry of the lineshape (\ref{eq:VdcLor}) depends on the
anisotropy of the torque $\vtau_{\rm eq}$ and on the angle $\phi/2$
between the principal axes and the direction $\vn$ of the
magnetization of the fixed layer. In the experiment of Ref.\
\onlinecite{kn:sankey2006} the main contribution to $\vtau_{\rm eq}$
comes from the large magnetic field used to align the free layer
magnetization perpendicular to $\vn$. This contribution is isotropic,
which explains why no strongly asymmetric lineshapes were observed in Ref.\
\onlinecite{kn:sankey2006}. The experiment of Ref.\
\onlinecite{kn:tulapurkar2005} finds a significantly asymmetric
lineshape if the applied magnetic field is small,
the lineshapes becoming more symmetric at larger fields. 
Although this observation
appears consistent with
our theory, we should note that for Ref.\ \onlinecite{kn:tulapurkar2005}
the equilibrium torque $\vtau_{\rm eq}$
arising from the applied magnetic field and shape anisotropy alone has 
$\phi=0$ and, hence, cannot explain an asymmetric lineshape.
Reference \onlinecite{kn:tulapurkar2005} attributes the
asymmetric lineshape to the imaginary part $g_2$ of the mixing
conductance which, if large enough, provides an alternative 
(but approximately magnetic-field independent) mechanism for
an asymmetric lineshape, see the discussion below.

The relative contributions of the rectification and the spin
emission effects can be found by looking at the ratio of $m_0 \langle
J m_1\rangle/e$ and $\langle \dot{m}_2 m_1 \rangle \tilde
  \alpha_-$, {\em cf.}\ Eq.\ (\ref{eq:V}). 
For $\tilde \alpha_{\pm} \ll 1$ this ratio is
\begin{eqnarray}
  \frac{m_0 \langle
  J m_1\rangle}{e \langle \dot{m}_2 m_1 \rangle \tilde
  \alpha_-} &=& - 2
  \frac{\delta - \omega_- \sin \phi (\omega-\omega_0)/\omega_0}
  {\tilde \alpha_-(\omega_+ - \omega_- \cos \phi)}.~~
  \label{eq:ratio}
\end{eqnarray}
Since both terms in the numerator are of order 
$\delta$ near the ferromagnetic
resonance, whereas the denominator is of order $\tilde \alpha_-
\omega_0$, the ratio (\ref{eq:ratio}) is of order
$\delta/\alpha_- \omega_0$. This is of order unity if $\tilde \alpha_+$ and
$\tilde \alpha_-$ are comparable, which happens precisely if the
second term in Eq.\ (\ref{eq:alphadef}) is not small in comparison to 
the first. This, in turn, is the condition that spin emission gives a
significant contribution to the total damping. Hence, we conclude
that spin emission contributes significantly to the
measured dc voltage if and only if spin emission contributes
significantly to the damping. Since $\langle \dot{m}_2 m_1 \rangle$ is
symmetric around $\omega = \omega_0$, {\em cf.}\ Eq.\ (\ref{eq:m2m1})
above, spin emission contributes to the symmetric part of the
lineshape only. The antisymmetric part is due to the rectification of
the applied ac current only.

In our calculations we have neglected the imaginary parts $g_2$ and
$G_2$ of the mixing conductance because in metallic junctions they
 are known to be
numerically small in comparison to the real parts $g_1$ and
$G_1$. Inclusion of $g_2$ and $G_2$ leads to a small modification of
the resonance frequency, because $g_2$ and $G_2$ change the
gyromagnetic ratio $\gamma$ of the free ferromagnetic 
layer.\cite{kn:tserkovnyak2002} With corrections to
first order in $g_2/g_1$ only, the resonance frequency becomes
\begin{eqnarray}
  \omega_0^2 &=& (\omega_+^2 - \omega_-^2) \\ && \mbox{} \times
  \left[1 - \frac{4 \tilde \alpha_- g_2(g_1 G_{1+} + 2 G_1
  G_{1+} - G_1 G_{+})}{g_1(G_1 G_+ + g_1 G_+ - g_1  G_{1+})}
  \right].\nonumber 
\end{eqnarray}
More importantly, with
nonzero $g_2$ and $G_2$, there is a finite asymmetry 
in the lineshape even in the absence of magnetic
anisotropy in the free layer,\cite{kn:tulapurkar2005}
\begin{eqnarray}
  \delta' &=& \frac{2 \omega_0 [\omega_- \sin \phi - z (\omega_+ + \omega_- \cos
    \phi)]}{2 \omega_+ \tilde \alpha_+
  + \omega_+ \tilde \alpha_- + \omega_- \tilde \alpha_- \cos \phi
  - 2 z \tilde \alpha_- \omega_- \sin \phi }, \nonumber \\
\end{eqnarray}
with 
\begin{equation}
  z= 
 \frac{G_1^2 g_{1+} g_2 G_- - 2 g_1^2
  G_{1+} G_2 g_-}{g_1 G_1 (g_1 + G_1) g_{1+} G_-}.
\end{equation}
Again, our results are valid up to first order in $g_2/g_1$ and
$G_2/G_1$ only.

We have also analyzed the case that the equilibrium
angle between the fixed layer magnetization $\vn$ and the free layer
magnetization $\vm$ is not 90 degrees. While this complicates the 
detailed expression for $V_{\rm dc}(\omega)$ (to the extent that 
it cannot be reported here), it does not change our
qualitative conclusions that (i) spin emission and rectification of
the applied ac current have comparable contributions to the generated
dc voltage if the free layer is thin enough that spin emission gives a 
sizable enhancement of the damping and (ii) the asymmetry of $V_{\rm
  dc}(\omega)$ around the resonance frequency $\omega_0$ is small in
the ratios $\omega_-/\omega_+$ or $g_2/g_1$. The former ratio is small if the 
applied magnetic field is large enough to saturate the free
ferromagnet, whereas the latter 
ratio $g_2/g_1$ is known to be numerically small for metallic
junctions
(of order $0.1$ or less, see 
Refs.\ \onlinecite{kn:xia2002,kn:zwierzycki2005,kn:zimmler2004}) .

\section{Conclusion}
\label{sec:5}

In this article we have presented a microscopic theory for the
spin-torque driven ferromagnetic resonance in
ferromagnet--normal-metal--ferromagnet trilayers. Our theory is
inspired by the experiments of Refs.\ \onlinecite{kn:tulapurkar2005}
and \onlinecite{kn:sankey2006}. In these experiments, an alternating
current is used to drive the ferromagnetic resonance, while a
generated dc voltage is used to detect the resonance. 

In addition to providing theoretical expressions for the width and
asymmetry of the resonance, we are able to determine the relative
magnitude of two physical mechanisms that contribute to the dc
voltage: rectification of the applied ac current and rectification of
the spin currents emitted by the precessing ferromagnet. Both
contributions are of similar magnitude for the thin ferromagnetic
films used in the experiments. The presence of two mechanisms to
generate a direct response to periodic driving, rather than one, sets
this class of magnetic devices apart
from their semiconductor counterparts.

A direct experimental probe of the two contributions to the dc voltage
is to compare the dc voltage observed in spin-torque driven
ferromagnetic resonance with the dc voltage generated in conventional
magnetic-field driven ferromagnetic resonance in the same device. 
The latter follows from
rectification of emitted spin currents only. Since spin emission gives
a symmetric line shape around the resonance frequency
$\omega = \omega_0$ there should be a
clear difference between the two methods to excite ferromagnetic
resonance. A comparison of the magnitudes of both
contributions would require a calibration of the amplitude at which
the magnetization precesses. 
This can be achieved through a simultaneous measurement of the dc
resistance of the device, which depends on the precession amplitude
through the giant magnetoresistance effect. 

\acknowledgments

We thank D.\ Ralph and J.\ Sankey for stimulating discussions. 
This work was supported
by the Cornell Center for Materials research under NSF grant no.\
DMR 0520404, the Cornell Center for Nanoscale Systems under NSF 
grant no.\ EEC-0117770, by the NSF
under grant no.\ DMR 0334499, and by the Packard Foundation.
Upon completion of this manuscript we learned of a work by A.~A.\
Kovalev, G.\ E.\ W.\ Bauer, and A.\ Brataas with similar conclusions
about spin-torque driven ferromagnetic resonance.\cite{kn:kovalev2006}
We thank Alex Kovalev for sending us a preprint of this work.


\begin{thebibliography}{24}
\bibitem{endnote15}
{A finite resistance of the normal-metal spacer layer (but without spin relaxation) can be taken into account by making the replacements $G_{+} \to 2 G_N (G_N G_+ + G_+^2 - G_-^2)/((G_N+G_+)^2 - G_-^2)$, $G_{-} \to 2 G_N^2 G_{-}/((G_N+G_+)^2 - G_-^2)$, $G_1 \to G_N(G_1^2 + G_1 G_N + G_2^2)/((G_1+G_N)^2 + G_2^2)$, and $G_2 \to G_N^2 G_2/((G_1+G_N)^2 + G_2^2)$ in the equations below, where $G_N$ is the conductance of the normal metal (per spin direction). A finite resistance of the free ferromagnetic layer can be taken into account by making the replacements $1/g_{\delimiter "3222378 } \to 1/g_{\delimiter "3222378 } + 1/2 g_{F\delimiter "3222378 }$, $1/g_{\delimiter "3223379 } \to 1/g_{\delimiter "3223379 } + 1/2 g_{F\delimiter "3223379 }$, with $g_{\pm } = (g_{\delimiter "3222378 } \pm g_{\delimiter "3223379 })/2$, while keeping $g_1$ and $g_2$ unchanged. Here $g_{F\delimiter "3222378 }$ and $g_{F\delimiter "3223379 }$ are the majority and minority conductances of the ferromagnet. In all cases, contact resistances should be excluded from the conductances.}
\end{thebibliography}

\end{document}